%
%
%
%
%
%
%
%

\input amstex
\catcode`\^^J=10
\magnification=\magstep1
\documentstyle{amsppt}
\NoBlackBoxes
\pagewidth{6.4 truein}
\pageheight{8.1 truein}
\hcorrection{.37  truein}
\vcorrection{.1 truein}
\nologo
\TagsOnRight


\output={\plainoutput}

\footline={\hss\bf\folio\hss}
\headline={\hfill}
\def\ovarphi{{\varphi^{\kern-3.5pt \raise 2 pt \hbox{$\scriptscriptstyle 
0$}}}{} }

\def\og{{g^{\kern-4pt \raise 1.5 pt\hbox{$\scriptscriptstyle 0$}}}}

\def\hook{\mathbin{\raise2.5pt\hbox{\hbox{{\vbox{\hrule height.4pt 
width6pt depth0pt}}}\vrule height3pt width.4pt depth0pt}\,}}

%
%

\document 

{\nopagenumbers
\line{}

\vskip 1.5 true in
\centerline{\bf ASYMPTOTIC CONSERVATION LAWS IN FIELD THEORY } 

\vskip 60pt

\centerline{\smc Ian M. Anderson${}^{(1)}$ \quad 
            and  \quad Charles G. Torre${}^{(2)}$    }
\bigskip
\line{$(1)$   Department  of  Mathematics,   Utah   State 
University, Logan, Utah 84322--3900.\hfill} 
\medskip
\line{$(2)$   Department of Physics, Utah State University, 
Logan, Utah 84322--4415. \hfill}

\vskip 60pt 
\baselineskip =20pt

\subheading{Abstract} A new, general, field theoretic approach to the
derivation of asymptotic
conservation laws is presented. In this approach asymptotic conservation laws
are constructed directly {}from the
field 
equations according to a universal prescription which does not rely upon the
existence of Noether identities or 
any Lagrangian or Hamiltonian formalisms.  The resulting general expressions
of the conservation laws enjoy important
invariance properties and synthesize all known asymptotic conservation laws,
such as the ADM 
energy in general relativity.

\medskip
\noindent
{\bf PACS:\quad} 3.50.-z, 11.30.-j, 4.20.-q

\vfill

\newpage}

\pageno=1
\headline={\hfill}
\baselineskip =20pt

In  
nonlinear gauge theories such as Yang-Mills  theory and general 
relativity, conserved quantities such as charge and energy-momentum
are computed {}from the limiting values of 
2-dimensional surface integrals in asymptotic regions.
Such  {\it asymptotic  conservation laws\/} are most often 
derived by one of two, 
rather distinct,  methods. One method, applicable to gauge theories such as
Yang-Mills and general relativity, relies upon the construction
of identically conserved currents furnished by Noether's theorem \cite{8},
\cite{10} and subsequent extraction of
a ``super-potential'' to define a conserved surface integral. 
Unfortunately, there is no general
field-theoretic criterion to select the appropriate
current:  for {\it any \/} field
theory there are infinitely many currents that can be
expressed as the divergence of a skew-symmetric super-potential,
modulo the field equations.  This
method of finding asymptotic conservation laws is thus somewhat {\it ad
hoc}.  An alternative approach to finding asymptotic
conservation laws in gauge theories is based upon the Hamiltonian
formalism.  In this approach, asymptotic conservation laws arise as
surface term 
contributions to symmetry generators \cite{4}, \cite{5}, \cite{10},
\cite{11}.  The
Hamiltonian approach to finding asymptotic conservation laws
lacks the {\it ad hoc} flavor of
the super-potential formalism   but is somewhat
indirect:   to construct asymptotic conservation laws using this method  one
must have the Hamiltonian formalism well in hand and 
 one must know {\it a priori} the general form of the
putative symmetry generator in order to find the appropriate surface
integral.

The purpose of this letter is to describe a new, general
construction of asymptotic conservation laws  for classical field 
theories which provides a viable alternative to existing methods. Our
construction is based upon a remarkable new spacetime differential form
$\Psi$,
which we derive for {\it any} system of field equations.  We show, by
virtue of an identity  involving the exterior derivative 
$D\Psi$  and the  linearized  field equations, that $\Psi$ defines asymptotic
conservation laws (when they
exist) for any
field theory.  Thus we are able to establish that asymptotic conservation laws
for field theories can be viewed as arising {}from
(asymptotically) closed differential forms canonically associated to the field
equations.  We also show that $\Psi$ possesses a number of attractive field
theoretical properties. In particular, this differential form is constructed
directly {}from the
field 
equations according to a universal prescription which does not rely upon the
existence of Noether identities or 
any Lagrangian or Hamiltonian formalisms. The form enjoys important
invariance properties and synthesizes all known asymptotic conservation laws,
for example, it is easily seen to
reduce to the ADM 
energy in general relativity.  We emphasize that in obtaining these results
no {\it a priori} assumptions are made concerning the space of gauge
symmetries of the theory.  The symmetries required for the existence of
asymptotic conservation laws are automatically derived as a consequence of
our general formalism.  Thus our formalism contains a number of distinct
field theoretical advantages.  With regards to specific applications, we are
able to extend existing results on asymptotic conservation laws in a number
of ways.  For example, we 
obtain covariant expressions for asymptotic conservation laws which can be
used to
generalize the conservation laws in asymptotically flat general
relativity and Yang-Mills theory to other asymptotic structures.  We also
obtain a formula which generalizes the asymptotic conservation laws of the
Einstein equations to arbitrary second-order metric theories.  To illustrate
this latter point, we consider a class of ``string-generated gravity models''
\cite{6}.  It follows immediately {}from our formalism that the
standard ADM formulas still describe conservation of energy, momentum, and
angular momentum for these theories, at least in the asymptotically flat
context.

To describe our results in more detail, let us fix an 
$n$-dimensional spacetime manifold $M$,  
with local coordinates $x^i$, $i=0$, 1, \dots, $n-1$  and let us label the 
totality of fields  for our classical theory by $\varphi^A$, $A= 1$, 
\dots, $N$. These fields are  subject to a system  of field equations 
which
we write as 
$$ 
     \Delta_B(\varphi^{A}, \varphi^A{}_{,i}, \varphi^A{}_{,ij}) = 0.
\tag1 
$$
We have postulated that the field equations are of second-order only 
for the sake of simplicity. The general theory which we outline here 
is developed in \cite{2} for field equations of arbitrary order. 
To address the problem of finding asymptotic conservation laws for (1), we 
begin by first broadening the usual notion of a local conservation  law.
We say that a {\it conservation law of degree $p$}  for (1) is a spacetime
$p$-form
$$
     \omega    =   \omega_{k_1\cdots    k_p}[x^i,    \varphi^A] 
     \, dx^{k_1}\wedge  \dots  \wedge dx^{k_p},
\tag2
$$
depending locally on the fields $\varphi^A$ 
and their  derivatives to some finite order, 
such that the exterior derivative 
$$ 
    D\omega= (D_k\omega_{k_1\cdots k_p})\, dx^k\wedge 
    dx^{k_1}\wedge   \dots  \wedge dx^{k_p},
\tag3									
$$
vanishes on all solutions to the field equations. In (3), 
$D_k$ is the usual total derivative operator. 
 When $p=n-1$, we may 
express (2) in the form 
$
     \omega = \varepsilon_{kk_1\cdots k_{n-1}} J^k
     dx^{k_1}\wedge \dots \wedge dx^{k_{n-1}}
$
in which case the vanishing of (3) coincides with the vanishing of
the divergence $D_k J^k$ of the  current $J^k$. Accordingly, we shall 
say that the $p$-form conservation law 
(2) is an {\it ordinary or classical conservation law\/}
in the case $p=n-1$ and a {\it lower-degree conservation law \/} when
$p < n-1$.

\par

Recently (\cite{2}, \cite{3}, \cite{7}, \cite{12}, \cite{14})
a number of methods have been developed for the systematic 
computation of all lower-degree conservation laws for  field 
equations such as (1).  
{\it The central premise of this note is 
that the techniques introduced in \cite{2} can be successfully adapted to 
the analysis of asymptotic conservation laws.} 
The principal ideas are as follows. If $\omega$ is a lower-degree 
conservation law for $\Delta_B=0$, then it is readily  checked that
the variation, $\delta_h\omega$,  of $\omega$ with respect to 
field variations $h^A= \delta \varphi^A$ is  closed by virtue of the
field equations $\Delta_B= 0$ and their formal linearization 
$\delta_h\Delta_B = 0$. Such  linearized conservation laws play a
pivotal role in the general theory of lower-degree conservation laws,
as presented in \cite{2} (see also \cite{13}). 
In \cite{2} it is  shown that every closed $p$-form $\omega[h]$,
depending  upon the fields $\varphi^A$ and their derivatives
and linearly on the field variations $h^A$ and their derivatives, can 
be   cast 
into a   {\it universal normal form\/} $\Psi_\rho[h]$, which depends upon
certain 
auxiliary fields $\rho$  that are subject to a set of algebraic and 
differential  constraints.  This normal form, which forms the basis for our
construction of asymptotic conservation laws, is obtained as follows.   
Details, generalizations, and further examples  of this construction
will appear in \cite{2}. 

To begin, it is  helpful to 
write the   formal linearization of the field equations (1) as
$$
     \delta_h\Delta_B=  \sigma{}_{AB}^{ij}h^A{}_{,ij}+
     \sigma{}_{AB}^i h^A{}_{,i} + \sigma{}_{AB} h^A.  
$$
We then define  a {\it  linear lower-degree conservation  law\/} for 
the field  equations (1) to be a  $p$-form $\omega[h]$, 
where $p<n-1$, of the type
$$
     \omega[h]= M_Ah^A + M_A^{i_1}h^A{}_{,i_1} +\dots + 
     M_A^{i_1i_2\cdots i_k} h^A{}_{,i_1i_2\cdots i_k},
\tag4
$$
which satisfies
$$
     D\omega[h]= \rho^B \delta_h\Delta_B + \rho^{Bi_1}D_{i_1}\delta_h
\Delta_B
     +\dots + \rho^{Bi_1i_2\cdots i_k}D_{i_1i_2\cdots 
i_k}\delta_h\Delta_B.
\tag5
$$
In equations (4) and (5)   the coefficients 
$M_A^{i_1i_2\cdots i_l}$ and $\rho^{Bi_1i_2\cdots i_l}$
are  spacetime $p$-forms and $(p+1)$-forms respectively, 
which depend on the fields $\varphi^A$ and their derivatives.
Equation (5) is  an identity in the field variations $h^A$ and their
derivatives but is still subject to the field equations $\Delta_B=0$.
The next step is to derive equations for these  multipliers $\rho$  
{}from the 
integrability condition $D^2\omega[h]=0$. It is 
a remarkable fact that the highest order multiplier 
$\rho^{Bi_1i_2\cdots i_k}$ is thus constrained by the  purely
algebraic condition 
$$
     \rho^{B(i_1i_2\cdots i_k}\sigma_{AB}^{hj}\wedge dx^{l)}= 0. 
\tag6
$$
We call this fundamental equation {\it the algebraic Spencer equation
for the linear conservation law $\omega[h]$ for the field equations 
(1)} \cite{15}.  For  the field equations that one typically considers, it is
not too
difficult to apply standard methods {}from tensor algebra to solve the
Spencer equation (6) \cite{2}.  The solutions to (6) often allow one to
simplify the
identity (5), by repeated ``integration by parts'', to the reduced form
$$
      D\omega[h]= \rho^B \delta_h\Delta_B.
\tag7
$$
We remark that for Lagrangian theories, full knowledge of the gauge 
symmetries of the theory  often  allows one to pass directly to the reduced
form (7). 

Let $\rho^{Bi} = dx^i \wedge\rho^B$ and 
$\mu^B_i = \dfrac{\partial\hfill}{dx^i} \hook \rho^B$.  
Assuming that the reduced equation (7) holds, we then have the following
complete characterization of  all  linear lower-degree conservation laws
of degree $p <n-1$ \cite{2}. 

\proclaim{Theorem (Classification of Linear Lower Degree Conservation Laws)} 

Let  $\Delta_B(\varphi^A,  \varphi^A{}_{,i}, \varphi^A{}_{,ij}) =0$  be  a 
system of second order field equations and suppose the linear 
$p$-form $\omega[h]$ satisfies the reduced equations (7). Then
\smallskip\noindent
{\rm [i]} the conservation law multiplier $\rho^B$ satisfies 
$$
\rho^{B(i}\sigma{}_{AB}^{jk)}   = 0,\qquad
\rho^{B(i}\sigma{}_{AB}^{j)} +    
     D_k\bigl[\rho^{B(i}\sigma{}_{AB}^{jk)} \bigr]
     =0, \qquad
   \rho^{Bi}\sigma{}_{AB} + 
     D_j\bigl[\rho^{B(i}\sigma{}_{AB}^{j)}\bigr]
    =0;
\tag8a,b,c
$$
{\rm [ii]}  the $p$-form 
$$
     \Psi_\rho[h]
    =  \frac1{q}    \, h^A\bigl[ \mu^B_i \sigma{}_{AB}^i -  
       \frac{2}{q+1}\, D_j( \mu^B_i \sigma{}_{AB}^{ij}\bigr)]
    +  \frac{2}{q+1}\, h^A_j\bigl[\, \mu^B_i\sigma{}_{AB}^{ij}\bigr],
\tag9
$$
where $q=n-p$, is closed by virtue of the equations $\Delta_B=0$, 
$\delta_h \Delta_B = 0$ and (8); and  

\smallskip\noindent
{\rm [iii]} every linear $p$-form $\omega[h]$ satisfying (7) differs {}from 
$\Psi_\rho[h]$ by an exact linear $(p-1)$-form.
\endproclaim
\noindent
We remark that this theorem can be readily generalized to the case where (5)
holds, rather than (7). 

To use this theorem for computing asymptotic
conservation laws we proceed as follows.  Suppose the 
spacetime $M$ is non-compact and admits an asymptotic region which 
is diffeomorphic to $\text{\bf R} \times C'$, where $C'$ is the 
complement of a compact set in $\text{\bf R}^{n-1}$.
Fix local   coordinates $(t, x^1, \dots, x^{n-1})$ 
in  the asymptotic region.
We consider a fixed solution $\ovarphi^A$  to the field equations 
and then, given a second solution $\varphi^A$, 
we set $h^A =\varphi^A- \ovarphi^A$. 
Let  $\omega[\ovarphi,h]$ be a spacetime $(n-2)$-form  
depending locally on the fields
$\ovarphi^A$ and $h^A$ and their derivatives.
We call  $\omega$ {\it an asymptotic conservation law
for the field equations $\Delta_B =0$ relative to the background
$\ovarphi$}  if, whenever $h^A$  satisfies the appropriate asymptotic
decay  condition as $r = \sqrt{(x^1)^2 + (x^2)^2 +\dots +(x^{n-1})^2} \to 
\infty$, the $p$-form $\omega[\ovarphi,h]$  satisfies 
$
     \omega\sim \text{O}(1)
$
and
$  
     D\omega\sim \text{O}(1/r).
$
Under these conditions
the limit 
$
    \dsize \lim_{r\to\infty} \int_{S_{(r,t_0)}} \omega[\ovarphi,h],
$
where $S_{(r,t)}$ is an $(n-2)$-dimensional sphere of radius $r$ in the 
$t=t_0$ hypersurface,  exists and is independent of $t_0$.  We therefore 
can deduce  {}from our classification theorem  that  if the asymptotic boundary
conditions are such that

\medskip

\itemitem{[i]} \qquad $\Psi_{\rho}[h] \sim \text{O}(1)$,

\medskip

\itemitem{[ii]} the equations $\delta_h \Delta_B(\ovarphi) = 0$ 
hold asymptotically to an order such that 
$\rho^B \delta_h \Delta_B(\ovarphi) \sim \text{O}(1/r)$, and
\medskip

\itemitem{[iii]}the conservation law multiplier equations (8) hold,
not exactly and not for all fields values, but only at $\ovarphi$ and
only to the appropriate asymptotic order, 

\medskip

\noindent
then {\it the normal form $\Psi_\rho[h]$ will be an asymptotic conservation 
law for the field equations $\Delta_B =0$ relative to  the  background
solution $\ovarphi$}. We note that the conditions [i], [ii], and [iii]  can 
be 
relaxed   so  long  as the relevant integrals arising {}from the application
of  Stokes Theorem to (7) vanish asymptotically. 

To  demonstrate the utility of our general theory  of  asymptotic 
conservation  laws, and to expose some of its novel features, we consider
the  vacuum Einstein  
field  equations. Here, as a consequence of (8a), the auxiliary fields
$\rho$  are
determined by a single vector field $X$ and the normal  form $\Psi_{\rho}[h]$ 
becomes 
$$
     \Psi_X[h]
     = \frac1 {16 \pi}
      \varepsilon_{ijhk} \bigl[ 
     -\frac23 (\nabla_u h_{rs})X_t \sigma^{ ti,rsuj} 
    + \frac13 h_{rs} (\nabla_u X_t) \sigma^{ ti,rsuj}\bigr]
      \, dx^h \wedge dx^k,
\tag10
$$
where  $\nabla$  is  the  covariant  derivative  defined  by  the 
Christoffel symbols of  $\og$ and  
$
     \sigma^{rs,ijhk}    =    
     {\partial     G^{rs}}/{\partial g_{ij,hk}}(g_0)
$
is the symbol of the Einstein  tensor. In expanded form, $\Psi_X[h]$  becomes 
$$
     \Psi_X[h]
     = \frac1{32 \pi}\varepsilon_{ijhk}\bigl[ h^{li}\nabla_lX^j 
     - \frac12h \nabla^iX^j -X_l\nabla^ih^{lj} 
     +X^i(\nabla_l h^{jl} - \nabla^jh)]\, dx^h\wedge dx^k,   
\tag11
$$
where  $h = \og^{ij} h_{ij}$, and indices are raised and 
lowed with respect to the background metric.  The differential form (11) 
recently appeared in 
\cite{10} (see equations (61), (63) and (75)),  where it was derived {}from 
the Noether identities and used in the study of black hole entropy.
 
\par

The spacetime form $\Psi_X[h]$ 
possesses the following desirable properties which reflect general features of
our formalism \cite{2}.

\smallskip

\itemitem{[1]}  The form $\Psi_X[h]$ is  closed by virtue of  the  the
Einstein equation for
the background $\og$, and the linearized Einstein equations for the
variation  $h$, provided that $X$ is Killing vector field of the background.
By using the classification theorem
above it can be shown that,  modulo  2-forms which are  exact  by 
virtue of these equations,  $\Psi_X[h]$ is the {\it only} linear lower-degree
conservation law for the vacuum Einstein equations.

\smallskip

\itemitem{[2]}  It is readily established that under the 
gauge  transformation 
$
     h_{ij} \to h_{ij} + \nabla_{(i}Y_{j)}
$
the form $\Psi_X[h]$  changes by an exterior derivative of a canonically 
defined 1-form. This implies that $\Psi_X[h]$ is suitably ``gauge invariant''.

\smallskip

\itemitem{[3]} The form $\Psi_X[h]$ is constructed directly {}from 
the  field  equations in a covariant fashion and with  no reference  made 
to  the  Bianchi 
identities  or  to  any Lagrangian   or  Hamiltonian.    Indeed, most 
properties  of  $\Psi_X[h]$ can be  inferred  directly 
{}{}from properties of  the symbol of the Einstein tensor. 

\itemitem{[4]}  If the vector field
$X$  is an asymptotic Killing  vector field for $\og$ and if $h = 
g - \og$ satisfies  appropriate decay conditions, then 
$\Psi_X[h]$  is always an asymptotic conservation law for the Einstein field
equations.  

\smallskip

\itemitem{[5]} In terms  of the (3+1) formalism, a lengthy but 
straightforward 
computation shows that  the pullback $\omega$ of $\Psi_X[h]$ to a leaf  of 
a foliation of spacetime by spacelike hypersurfaces becomes 
$$
\aligned 
     \omega=\frac{1}{16\pi} [
&    G^{abcd}(X^\perp \tilde\nabla_b h_{cd}-
     h_{cd}\tilde\nabla_bX^\perp)+ 2X^b\delta_h\pi_b^a
     -X^ah_{cd}\pi^{cd} 
\\ 
& -\tilde\nabla_b(h^a_{\perp}  X^b - h^b_{\perp}X^a) ]\,   d^2S_a .
\endaligned
\tag12
$$
In this equation  
(i) $X^\perp$ and $X^b$  are the normal and tangential components
 of a Killing vector  with respect to the spacelike
 hypersurface and $h_\perp^a$ is the normal-tangential  component 
of $h^{ij}$; (ii) the derivative operator $\tilde \nabla_b$ is 
compatible with the background metric 
$\gamma_{ab}$ induced on the 
 hypersurface; (iii)  we have defined
$$
     G^{abcd}={1\over 2}\gamma^{1/2}
     (\gamma^{ac}\gamma^{bd}+\gamma^{ad}\gamma^{bc}
     -2\gamma^{ab}\gamma^{cd});
$$
and (iv) $\pi^a_b= \gamma_{bc}\pi^{ac}$, where $\pi^{ac}$ is the
canonical field momentum. With asymptotically flat or 
asymptotically anti-de Sitter boundary conditions the 
integral of (12) over the sphere at 
infinity  coincides with 
the standard ADM-type formulas (\cite{4}, \cite{9}, \cite{11}) 
for the asymptotic conservation laws in this context.

\smallskip

Thus our construction is general enough to capture the usual conservation laws
in general relativity.  Moreover, $\Psi_X$ provides a means of generalization
of these conservation laws to other
asymptotic structures.  In particular, the presence  of the term $X^ah_{cd} 
\pi^{cd}$ in (12) coupling $h_{ab}$ to the canonical momentum of the
background,
which is dictated by our  derivation  of 
$\omega$  {}from  a covariant spacetime  2-form $\Psi_X[h]$, has to our
knowledge   not
appeared  in previous ADM-type  formulas.  While  this 
term  does  not contribute to the surface integrals  at  infinity     
for the asymptotically flat and asymptotically anti-de Sitter 
boundary  conditions,  we  believe that  its  inclusion  should  be 
neccessary for other asymptotic  conditions.

Let us briefly describe other applications of our formalism.  First of all,
the classification theorem we have derived for linear lower-degree
conservation laws is used in \cite{2} to classify all conservation laws for
the Einstein equations, Yang-Mills equations, and similar equations.  With
regards to asymptotic conservation laws, in Yang-Mills theory our
classification theorem leads to
the standard surface integral formulas in the asymptotically flat
context \cite{2}. Work is in progress to investigate Yang-Mills conservation
laws in
the presence of other asymptotic conditions.   As another application, we
consider the asymptotic conservation laws for any covariant metric theory
described by second-order field equations.  We can show that the normal form
(10) generalizes without change to all such theories (provided, of course,
that one uses the symbol appropriate to the field equations of interest).  To
illustrate this point, we consider the ``string-generated gravity
models'', which are based upon the Lovelock Lagrangian density \cite{6}. 
Using the modified symbol of the string-corrected Einstein
equations, it is
straightforward to compute the asymptotic conservation laws for
string generated gravity in the asymptotically flat context {}from (10). 
Because the string-corrected symbol reduces to that of the Einstein equations
on a flat
background, it follows immediatedly that the conservation laws (11), (12)
still apply in the presence of the corrections to general relativity
suggested by string theory.

In summary, the theory of  linearized conservation
laws leads  to a new, efficient, systematic, 
and covariant  method for   
obtaining asymptotic conservation laws in field theory.  We expect
that this approach will prove useful for a  wide range  of field
theories and asymptotic boundary conditions
which have  been heretofore  unexplored, and we anticipate that the
spacetime form $\Psi_\rho[h]$ will be a valuable tool in the further
study of conservation laws in field theory.

\medskip

\subheading{Acknowledgements} This research was supported, in part, by 
grants DMS94--03783 and PHY96-00616 {}from the National Science 
Foundation.  Both authors gratefully acknowledge the hospitality
of  the Centre de Recherches Math\'ematiques at the  Universit\'e 
de Montr\'eal  where this work was initiated.

\end